# Wide-Band Nano-Imaging of Plasmon Dispersion and Hotspots in Quasi-Free-Standing Epitaxial Graphene

*William S. Hart[+], Vishal Panchal[^], Christos Melios[^], Włodek Strupiński[†,‡], Olga Kazakova[^], Chris C. Phillips[+,*]*

[+] Experimental Solid State Physics Group, Physics Department, Imperial College London, London, SW7 2AZ, U.K.

[^] National Physical Laboratory, Hampton Road, Teddington, TW11 0LW, U.K.

[†] Institute of Electronic Materials Technology, Wolczynska 133, 01-919 Warsaw, Poland.

[‡] Faculty of Physics, Warsaw University of Technology, Koszykowa 75, 00-662 Warsaw, Poland.

ABSTRACT: We report observation of graphene plasmon interference fringes across a wide spectral range using a scattering scanning near-field optical microscope (s-SNOM) that employs a widely tunable bank of quantum cascade lasers. We use plasmon interference to measure the dispersion curve of graphene plasmons over more than an order of magnitude of plasmon wavelength, from $\lambda_{sp}$ ~140 to ~1700 nm, and extract the electron Fermi energy of 298±4 meV for hydrogen-intercalated single layer epitaxial graphene on SiC. Furthermore, we demonstrate the appearance of wavelength tuneable graphene plasmon reflection "hotspots" at single-layer/bi-layer interfaces. This work demonstrates the capability of wide-band nano-imaging to precisely measure the electrical properties of graphene and spatially control plasmon reflection focusing.

The Fermi energy, $E_F$, of graphene plays a central role in many of its applications, particularly in plasmonic devices[1–4]. Typical methods for measuring $E_F$ are the Hall effect, Kelvin probe force microscopy (KPFM), vector decomposition of Raman spectra and angle-resolved photoemission spectroscopy (ARPES). However, each method has its own set of advantages and drawbacks. For example, Hall effect measurements offer straightforward and direct measurement of the carrier density, which can then be used to calculate the Fermi energy for single-layer graphene (1LG) using $E_F^{1LG} = v_F \hbar \sqrt{(\pi n)}$, where $v_F$ is the Fermi velocity, $\hbar$ is reduced Planck's constant and $n$ is the carrier density of graphene. For bi-layer graphene (2LG), the Fermi energy is given by $E_F^{2LG} = \pi \hbar^2 n / 2m^*$, where $m^*$ is the effective mass of the charge carriers. However, the Hall effect method typically requires fabrication of Hall bar devices with the typical sizes ranging from hundreds of nanometers to several millimeters[5,6], or can also be performed in the van der Pauw geometry on continuous films. In both cases, it leads to averaging of the Hall voltage measurement over the large sample area and can therefore give a misrepresented $E_F$ for non-uniform graphene samples. KPFM is often used to map the local work function of graphene, which in turn can be used to estimate $E_F$. However, KPFM is extremely sensitive to surface contamination as well as substrate and environmental doping[7,8], and estimation of the sample's work function requires meticulous calibration of the KPFM probe[9]. The vector decomposition of the Raman G and 2D modes can also be used to determine the carrier density of graphene, but this method is currently limited to p-type 1LG samples[10]. Moreover, the spatial resolution of Raman mapping is limited by the relatively large spot size of the laser (100's of nanometers), making it unsuitable for samples with non-uniform layer thicknesses. Finally, nano-ARPES offers direct measurement of $E_F$,[11] but the technique is not widely available and requires ultrahigh vacuum conditions that significantly changes the doping of graphene, thus making it difficult to form direct comparisons to ambient conditions.

In 2012, Chen et al.[3] demonstrated that the wavelength of graphene plasmons can be measured by analysing the surface plasmon interference fringes that appear in images obtained by scattering-type scanning near-field optical microscopy (s-SNOM). Using a first order approximation of the plasmon dispersion relation, they inferred an approximate value for the Fermi energy of exfoliated graphene, $E_F \approx 400$ meV. Since then, many experiments have used this technique to study graphene and other layered 2D materials with applications in the study of chemical doping[1], grain boundary phenomena[12], graphene plasmonic nano-resonators[13,14], and hybrid systems with hyperbolic[15] or phonon polaritons[2]. However, the accuracy and energy range of these studies have so far been limited to a small number of discrete lines in a narrow spectral range, corresponding to free-space wavelengths of $\lambda_0$ ~9.2–10.6 μm that can be accessed by the $CO_2$ lasers typically used for s-SNOM measurements[3,16].

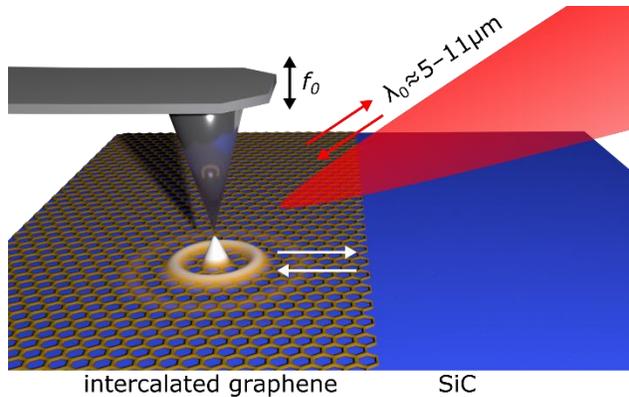

**Figure 1.** The probe with ~10 nm apex radius of curvature scatters surface plasmons (bright fringes) into graphene, overcoming the momentum mismatch between the plasmons and the light incident from the quantum cascade laser.

S-SNOM is a type of scanning probe microscopy that relies on "tapping mode" atomic force microscopy (AFM), i.e. with a vertical dither applied to the probe at its mechanical resonance frequency ($f_0 \sim 280$ kHz), which tracks the sample surface height. When the probe is in a close vicinity of the sample, the sharp platinum-coated tip radially scatters incident laser light into the graphene surface plasmon modes, which then propagate in the sample plane (Figure 1). When the plasmons encounter a defect, for example the edge of the 2D graphene layer, they are partially reflected back and produce a standing wave interference pattern, whose fringes have a spatial periodicity given by half the surface plasmon wavelength, $\lambda_{sp}/2$.[3] The amount of energy subsequently scattered out of the plasmon modes and back into the s-SNOM detection system is proportional to the field intensity at the probe, allowing the fringes to be mapped out in the s-SNOM image. The backscattered signal is demodulated at the second harmonic of $f_0$. This eliminates signals due to background scattered light, and makes the measurement sensitive only to light scattered by its interaction with the near-field intensity in the nanoscale region between the probe apex and the sample[17]. Here, we pair s-SNOM with a radiation source consisting of a bank of widely tuneable quantum cascade lasers (QCL; MIRcat, Daylight Solutions). This allows us ample capability to tune through a free-space wavelength range $\lambda_0$ = 8.93 to 10.43 μm, in a way that enables the full graphene plasmon dispersion curve to be obtained that is much broader than previously reported.

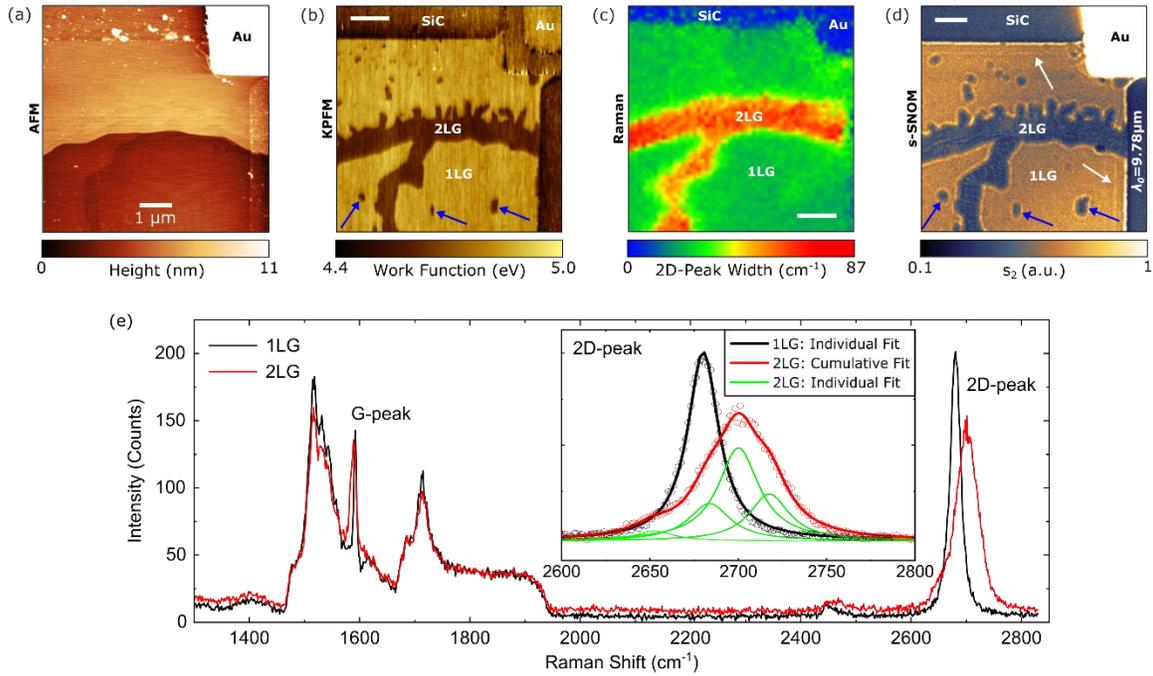

**Figure 2.** (a) Topography, (b) work function, (c) Raman map of the 2D-peak width and (d) s-SNOM amplitude, $s_2$, image obtained at $\lambda_0$ = 9.78 μm, for intercalated epitaxial graphene on SiC. Blue arrows in (b) point to 2LG islands. White and blue arrows in (d) point to plasmons reflecting from 1LG/SiC and 1LG/2LG interfaces, respectively. (e) Raman spectra for single- (1LG) and bi-layer graphene (2LG) showing typical features observed in this type of graphene. The inset in (e) shows the zoom-in of the 2D-peak with the characteristic single Lorentzian fitted to 1LG and four Lorentzians fitted to 2LG. All scale bars are 1 μm.

The graphene used in the present study was grown on semi-insulating SiC substrates in a hot-wall reactor (Aixtron VP508). First, a complete interfacial layer (IFL) was grown under an argon laminar flow at 1600 °C. Following IFL growth, the sample was annealed in hydrogen-rich environment. The hydrogen penetrates between the IFL and substrate, forming Si-H bonds, transforming the IFL to quasi-free-standing (QFS) 1LG. For more details, see Ref. 11. As it is not possible to entirely avoid nucleation growth in certain areas, e.g. near the edges of SiC terraces, such areas are covered with 2LG. QFS graphene on SiC produced with this growth method is typically p-type doped, which is attributed to the spontaneous polarisation of the hexagonal SiC[18,19].

Figure 2a shows the topography of the sample obtained with AFM, featuring height steps (2-6 nm) from the underlying SiC substrate, together with particulate remnants from the sample fabrication processes (small white spots). Figure 2b shows the work function ($\Phi$) image of the same area as Figure 2a. This was obtained by first calibrating the KPFM probe's work function ($\Phi_{Probe}$) against gold ($\Phi_{Au}$ = 4.82 eV) using $\Phi_{Probe} = \Phi_{Au} + eV_{SP}(Au)$, where $V_{SP}$ is the surface potential measured on gold, followed by applying $\Phi_{Sample} = \Phi_{Probe} - eV_{SP}(sample)$ to the surface potential image to obtain the work function of the sample[9]. Given that KPFM is only sensitive to changes in the surface potential, it is able to clearly distinguish the structure of 1LG, 2LG, SiC and Au due to their inherent differences in work function. For 1LG and 2LG, the work functions are $\Phi_{1LG}$ = 4.81 eV and $\Phi_{2LG}$ = 4.56 eV, respectively, and by assuming the intrinsic work function of graphene as $\Phi_0$ = 4.47 ± 0.05 eV from our previous work[8], we estimate the Fermi energy for 1LG as $E_F^{KPFM}$ = 340 ± 10 meV. This value compares well to Hall effect measurements in the van der Pauw geometry[20], $E_F^{Hall}$ = 333 ± 1 meV. The assignment of 2LG in the middle of the graphene sheet was also confirmed by mapping the Raman 2D-peak width, where it was determined as ~60

cm$^{-1}$ for 2LG versus ~20 cm$^{-1}$ for 1LG (Figure 2c)[21]. Furthermore, the 2D peak can be fitted by a single Lorentzian for 1LG and a broad peak with shoulders requiring four Lorentzians, which is characteristic for 2LG (Figure 2e). The s-SNOM amplitude, s$_2$, image in Figure 2d shows a strong mid-IR signal from the 1LG and reveals plasmon interference fringes at the 1LG/SiC and 1LG/2LG interfaces. The s-SNOM amplitude image shows clear interference fringes emanating from the right and top of the image (as indicated by the white arrows) that arise from the strong plasmon reflection from the straight, lithographically defined edges of the graphene sheet (i.e. the 1LG/SiC interface). For homogeneous 1LG, the full dispersion relation of plasmons as a function of optical frequency, ω, is given by [22]:

$$\frac{\varepsilon}{\sqrt{\varepsilon_{sub} k_0^2 - k_{sp}^2}} + \frac{1}{\sqrt{k_0^2 - k_{sp}^2}} = -\frac{4\pi\sigma(\omega)}{\omega} \qquad (1)$$

where $\varepsilon_{sub}$ is the substrate dielectric function, $k_0 \equiv \frac{2\pi}{\lambda_0} = \omega/c$ is the free space wavevector, $k_{sp} \equiv 2\pi/\lambda_{sp}$ is the in-plane plasmon wavevector and $\sigma(\omega)$ is the graphene conductivity. Under the random phase approximation, and for excitation energies sufficiently below $E_F$ so as to avoid interband transitions ($\hbar\omega \ll 2E_F$), the graphene conductivity depends linearly on the Fermi energy $E_F$ according to [22]:

$$\sigma(\omega) = \frac{e^2 E_F}{\pi\hbar^2} \frac{i}{\omega + i\tau^{-1}} \qquad (2)$$

where $e$ is the electronic charge and $\tau$ is the relaxation time of the charge carriers in the graphene.

By smoothly tuning the QCL through a wide range from $\lambda_0$ = 8.93 to 10.43 μm, in steps of 100 nm, we map out the variation of the fringe spacing with high spectral resolution. Figures 3(a) and 3(b) show maps of the plasmon interference fringes at $\lambda_0$ = 10.03 and 9.83 μm, respectively. Obtaining maps throughout the specified range of the QCL, we determine the graphene plasmon dispersion over more than a decade in the surface plasmon wavelength, $\lambda_{sp}$ = 140–1700 nm (Figure 3c). As expected, the fringe spacing, and so the plasmon wavelength, increases with increasing $\lambda_0$. Despite the larger error bar for the final data point, which stems from the poorer contrast of the peaks and troughs of the plasmon standing wave, we can extract the precise value of $E_F$ by combining equations (1) and (2) and fitting the experimental data with an error-weighted numerical minimisation. The resulting fitted dispersion relation yields $E_F^{SNOM} = 298 \pm 4$ meV (Figure 3c). The larger estimation of $E_F$ from the Hall effect is attributed to the averaging from the presence of a small quantity of 2LG, which typically exhibits a higher carrier density than 1LG[21], whereas for the work function measurements, the method generally exhibits higher uncertainty due to the low signal-to-noise ratio of surface potential measurements using frequency-modulated KPFM[9].

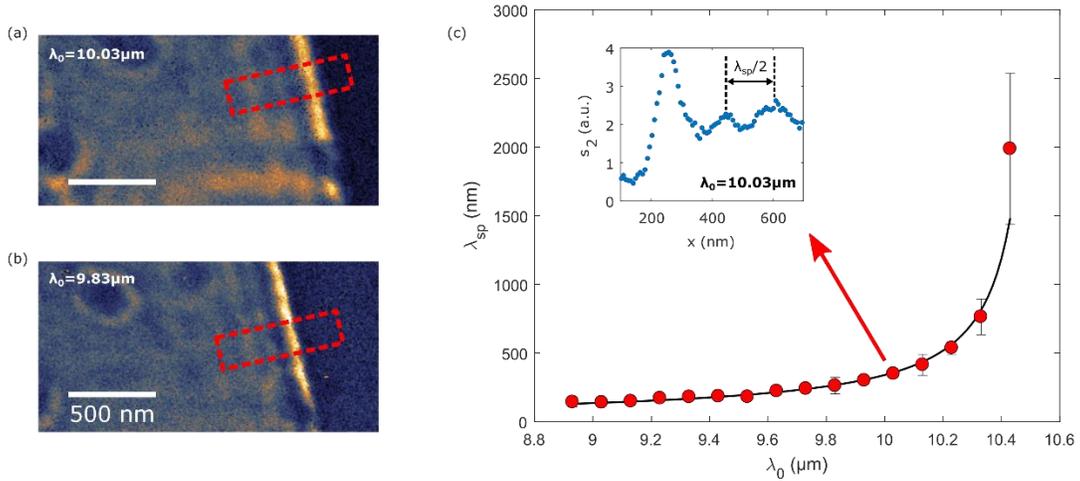

**Figure 3.** Plasmon interference fringes at (a) $\lambda_0$ = 10.03 μm and (b) $\lambda_0$ = 9.83 μm. (c) Calculated plasmon dispersion relation for graphene on SiC (black line) numerically fitted to the experimentally extracted plasmon wavelengths (red points), yielding a graphene Fermi energy $E_F = 298 \pm 4$ meV. The inset in (c) shows the line profile extracted from the red dashed rectangle in (a).

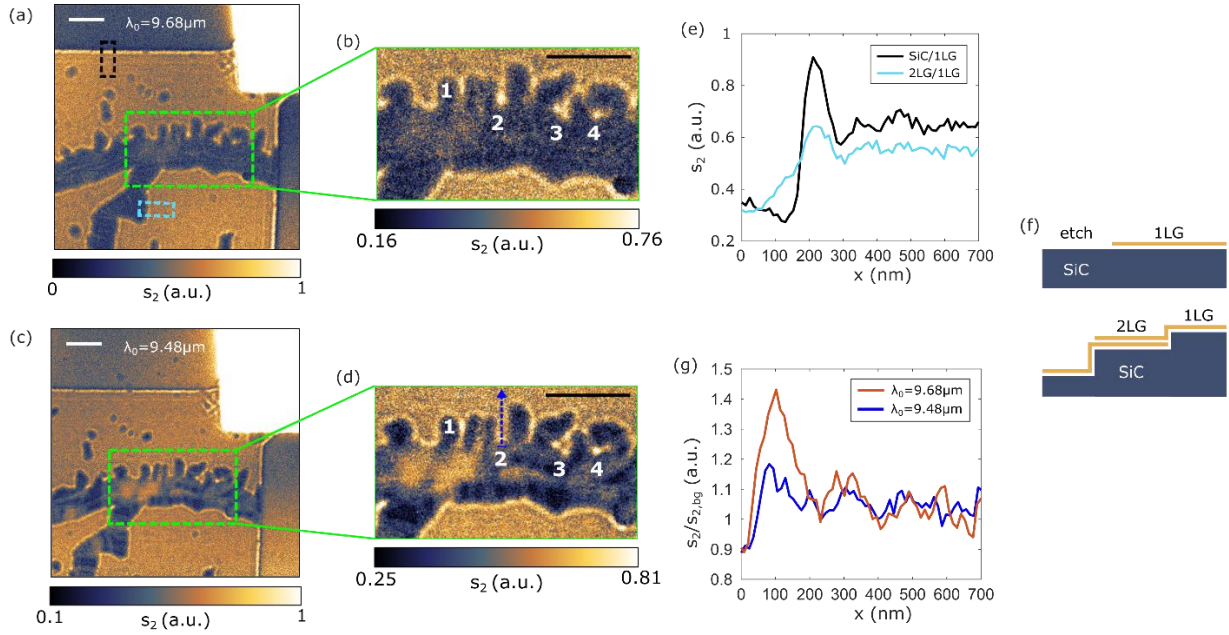

**Figure 4.** s-SNOM amplitude, $s_2$, images showing graphene plasmon interference at various interfaces at (a) $\lambda_0 = 9.68$ μm and (c) at $\lambda_0 = 9.48$ μm. (b, d) Magnified s-SNOM amplitude images showing constructive plasmon interference "hotspots" near 2LG "ribbons". (e) s-SNOM amplitude line profiles along the black and cyan regions indicated in (a) showing the relative strengths of plasmon reflections from SiC/1LG and 2LG/1LG interfaces, respectively, as illustrated in (f). (g) Line profiles across "hotspot" site 2 as indicated by the blue dashed arrow in (d) showing s-SNOM amplitude normalised to graphene background, $s_{2,bg}$. All scale bars are 1 μm.

Additional fringes are also present from plasmons scattering at the interfaces between 1LG and 2LG (Figures 4(a) and 4(c), obtained at $\lambda_0 = 9.68$ μm and $\lambda_0 = 9.48$ μm, respectively). These reflections are generally weaker than those at the edges of the 1LG sheets (i.e. from 1LG/SiC interface). Chen *et al.* reported on plasmons in graphene reflecting purely from the terrace step edges in SiC[16]. In our case, although there is an additional complication of 2LG also being present at step edges, we believe the latter is not the primary cause, given that we also observe plasmon reflections from smaller 2LG islands (blue arrows in Figure 2d), where the step height is ~0.35 nm (Figure 2a). The decrease in surface height from 1LG to the islands is consistent with epitaxial growth of 2LG rather than a hole in 1LG, which would show as higher. Furthermore, the magnitudes of the interference fringes around these islands are congruent with the fringes associated with the Raman-confirmed 1LG/2LG interfaces (Figure 2c), i.e. much smaller than those at the 1LG/SiC interface. In this case, the plasmon reflection occurs due to the change in carrier density across the 1LG/2LG interface (Figure 4e), as revealed by Fei *et al.*, where they apply a back gate voltage to vary the Fermi energy of single-, bi- and tri-layer graphene[23].

Finally, we observe a number of plasmon focusing "hotspots" along the edges of the 2LG "strips", whose relative intensities depend strongly on the graphene plasmon wavelength (magnified in Figures 4b and 4d). These strips of 2LG are known to form, as in this instance, along the terraces in the SiC substrate (as represented by the schematic in Figure 4f). When the laser is tuned from $\lambda_0 = 9.68$ μm (Figure 4b) to $\lambda_0 = 9.48$ μm (Figure 4d), the "hotspot" at position 2 entirely disappears, as seen in the $s_2$ line profiles in Figure 4g that are normalised relative to the background $s_{2,bg}$ of graphene away from the fringes. The "hotspot" at position 3 splits up into three separate "hotspots", and positions 1 and 4 increase in intensity relative to the graphene background. This occurs because the longer plasmon wavelength of Figure 4b means that the fringes are broader, so what would otherwise be three distinct "hotspots" blur into one.

Through the combination of scattering scanning near-field optical microscopy and a widely tunable QCL, our experiments comprehensively map the dispersion of graphene plasmons across more than an order of magnitude range of the plasmon wavelength. This wide coverage allows for precise extraction of the graphene Fermi energy entirely using this optical method. Our study establishes mid-infrared nano-imaging as a method for quantitative determination of the local electronic properties of graphene in ambient air and without the need for electrical contacts or any other specific sample preparation. Further, we have demonstrated wavelength tuneable "hotspots" arising from constructive interference of graphene plasmons with their reflections from interfaces between 1LG and 2LG. The spectral and spatial nature of these hotspots could be designed via manipulation of 2LG geometry, for use in enhanced chemical analysis, environmental monitoring, and plasmonic nanoantennas for boosting the sensitivity of fluorescence microscopy and vibrational spectroscopy.

AUTHOR INFORMATION

**Corresponding Author**

* E-mail: chris.phillips@imperial.ac.uk

**Author Contributions**

The manuscript was written through contributions of all authors. All authors have given approval to the final version of the manuscript.

**Funding Sources**

The work at Imperial College was sponsored by the Engineering and Physical Science Research Council (EP/K029398/1). The work at the National Physical Laboratory was supported by Graphene Flagship, 16NRM01 GRACE and the Department for Business, Energy and Industrial Strategy though NMS funding.

ACKNOWLEDGMENT

The authors thank Michael Winters for the fabrication of van der Pauw structures.